\documentclass[namedreferences,hyperref,optionalrh]{spr-sola}
\usepackage{graphicx}        
\usepackage{color}
\usepackage[utf8]{inputenc}
\usepackage{float}
\usepackage{geometry}
\usepackage{amsmath}                            

\usepackage{hyperref}
\hypersetup{
    colorlinks=true,
    linkcolor=red,
    filecolor=magenta,      
    urlcolor=cyan,
    citecolor=blue,
}

\chardef\us=`\_

\begin{document}

\begin{frontmatter}
\title{Multi-lane type II radio bursts: Insights into shock propagation in the corona}

\author[addressref={aff1}]{\inits{N.}\fnm{Nadiya}~\snm{K.}\footnote{This work was carried out as a final-year master's thesis by the author at USO-PRL.}}
\author[addressref={aff2,aff3}]
{\inits{D.}\fnm{Divya}~\snm{Paliwal\footnote{equal contribution}}\orcid{0009-0001-7689-0084}}
\author[addressref=aff2, corref,email={anshu@prl.res.in}]{\inits{A.}\fnm{Anshu}~\snm{Kumari\footnote{corresponding author}}\orcid{0000-0001-5742-9033}}
\address[id=aff1]{Mahatma Gandhi University,  Priyadarsini Hills Post, University Campus Rd, Athirampuzha, Kerala 686560 }
\address[id=aff2]{Udaipur Solar Observatory, Physical Research Laboratory, Dewali, Badi Road, Udaipur-313 001, Rajasthan, India}
\address[id=aff3]{Indian Institute of Technology, Gandhinagar, Gujarat-382355, India}

\runningauthor{Nadiya et al.}
\runningtitle{Multi-lane type II radio bursts}

\begin{abstract}

Type II solar radio bursts are considered as the signatures of the coronal shocks. These bursts are generated from plasma waves excited by magnetohydrodynamic (MHD) shocks, and then converted into radio waves at the local plasma frequency and/or its harmonics. Hence, these bursts often have fundamental-harmonic (FH) and band-splitting (SB) structures, which provide insights into shock generation and propagation in the corona, hence, in turn, the corresponding coronal conditions. 
In the present study, we analysed an unusual multi-lane type II burst observed with ground-based solar radio spectrographs on May 29, 2024, between 14:24 and 14:43 UT. The start and end frequencies of the type II burst were 450 MHz and 25 MHz, respectively. By combining spectral information with radio imaging data, we found that radio waves were escaping from the corona via emissions from distinct shock regions. In addition, along with the traditional FH and SB, there were multi-lane structures in the type II bursts. Our analysis suggests complex, inhomogeneous shock dynamics near the leading edge (LE) of the coronal mass ejection (CME). This indicates that the plasma material compresses more strongly in these forefront regions. This was confirmed by radio imaging observations, which showed that the higher-frequency emission occurred at a higher altitude than the lower-frequency emission. Our results suggest that the shock geometry and plasma inhomogeneity play an important role in the generation of type II bursts, leading to traditional fundamental-harmonic split-band (FH-SB) pairs with additional splitting in the type II bands.

\end{abstract}
\end{frontmatter}

\section{Introduction}
\label{section-sec1} 

The Sun, which is the closest star to us, emits radiation in all parts of the electromagnetic spectrum, from X-rays to radio waves \citep{Aschwanden2002, stix2012sun, Alissandrakis2019}. 
Solar transients, including solar flares, coronal mass ejections (CMEs), prominence eruptions, and related phenomena such as radio burst emissions, release large amounts of magnetic energy and plasma material into the heliosphere and the interplanetary medium \citep{hudson1991solar, Richardson2010, temmer2010combined,2002A&A...385.1078K} during and after the eruption.
The Sun releases off both thermal and non-thermal radiation. Background emission is thermal, whereas solar radio bursts are non-thermal \citep{Wild1963, gopalswamy2005type, Carley2020}. Solar radio bursts are the result of electrons being accelerated to near relativistic velocities \citep{cane1984type, Morosan2021, Kumari2023b, 2026arXiv260322087K}.  There are five main classifications of solar radio bursts (I–V) \citep{Wild50a, SuzukiDulk85}that depend on how they manifest in dynamic spectra  \citep{white2007solar}. These bursts are connected to certain physical processes, notably plasma emission and/or gyromagnetic activity in the solar corona.

Coronal shocks are considered as the source of type II radio bursts \citep{Smerd1962,ne85,Morosan2019a,Kumari2023b,Morosan2022b}. They are also the longest-known shock signature. They come from plasma waves that are caused by magnetohydrodynamic (MHD) shocks \citep{2005JGRA..11012S07G, Anshu2017a, 2025arXiv251221846B}.  These waves are then changed into radio waves at the local plasma frequency and/or its harmonics. So, the emission process is considered as plasma emission \citep{Wild1972, Tapping1978}.  Type II radio bursts offer substantial insights into the kinematics, dynamics, and energetics of solar eruptions in proximity to the Sun, facilitating the comprehension of physical processes within the solar corona and interplanetary plasma  \citep{cane1984type, kaiser1998type}. These bursts often have fundamental-harmonic (FH) pairings and band-splitting (SB) structures \citep{Smerd1974} which help us understand how shocks move and what the local corona is like. A high-speed shock wave speeds up electrons in the plasma emission mechanism. At the local plasma frequency (fp) and its harmonics, these Langmuir waves turn into electromagnetic radio waves. In dynamic spectra, type II bursts usually seem like two narrow lanes that slowly and steadily change frequency. Type II bursts seldom exhibit characteristics resembling "herringbones (HB)" signifying particle acceleration at the shock front \citep{ca87, 2023RAA....23e5010A}.
Recent statistical studies of type II radio bursts have shown that in most cases, these bursts are associated with CME-driven shocks \citep{Kumari2023b}. However, a few studies indicate otherwise. Type II bursts can occur due to the shock generated by a flare blast wave \citep{magdalenic2012flare}, a Jet-Driven shock \citep{Maguire2021}, or an EUV eruption \citep{2023arXiv230511545M, 2025JApA...46...90K}.\\

Imaging observations have long indicated that type II radio emission can originate from multiple spatially separated locations along a shock front. \cite{1969SoPh...10..460K} provided one of the earliest imaging evidence at Culgoora that type II emission from a single flare-associated event could arise from multiple spatially separated sources, with their positions inconsistent with smooth radial density models.
Previously, \cite{2017SoPh..292..194L} had reported multiple branches within a type II burst.
More recently, \cite{2020A&A...639A..56J} demonstrated through radio triangulation that low frequency type II sources often propagate through dense non-radial coronal structures, that is type II bursts are generated at the regions of higher density than what is usually expected while employing 1D density models. \cite{refId0} identified multiple emission regions along a single CME-driven shock using LOFAR high-resolution imaging.
Of late, \cite{2025A&A...703A.271Z} showed that type II bursts can have multiple lanes, each originating from a different part of the shock, including the shock front and the flank. 

In the present study, we add a further case to this growing observational corpus. We studied an unusual multi-lane type II burst observed on May 29, 2024, and exhibited multiple lanes in addition to the classical FH and SB pairs. Analyzed it using ground-based solar radio spectrographs and Nançay Radioheliograph (NRH) imaging, with the aim of identifying the spatial origin of the individual emission lanes and connecting the spectral morphology to the local coronal conditions through which the CME-driven shock propagates.

This article is organized as follows: Section \ref{section-sec2} describes the multi-wavelength ground- and space-based observations of these events; Section \ref{section-sec3} discusses the data analysis methods and the results. We discuss the results and conclude the paper in Section \ref{section-sec4}.


\section{Observations}
\label{section-sec2}

There was an X1.4 class recorded with the Geostationary Operational Environmental Satellites \citep[GOES;][]{2022SpWea..2003044D}-16 on May 29, 2024. The associated NOAA active region (AR) was 13697, and the flare location was S20E66 \footnote{\url{https://www.lmsal.com/solarsoft/latest_events_archive/events_summary/2024/05/29/gev_20240529_1411/index.html}}. 
The flare onset occurred at 14:11 UT, peaked at 14:23, and lasted approximately 44 minutes. Several space- and ground-based instruments have recorded the associated phenomenon at multiple wavelengths, including radio, extreme ultraviolet (EUV), and white-light observations. There was a halo CME recorded with the Large Angle and Spectrometric Coronagraph (LASCO) on board the Solar and Heliospheric Observatory. The CME was first visible in the LASCO-C2 field of view (FOV) at 14:38 UT and was also later seen in the LASCO-C3 FOV. The coronagraphs Sun Earth Connection Coronal and Heliospheric Investigation (SECHHI) on board Solar Terrestrial Relations Observatory \citep[STEREO-A;][]{2008SSRv..136....5K} had provided another fov since it was a limb event (S20E80) for the spacecraft. STEREO-A/COR1 observed the CME for the first time at ~14:31 UT. Different extreme ultraviolet channels of the Atmospheric Imaging Assembly (AIA) instrument on board the Solar Dynamics Observatory \citep[SDO;][]{2012SoPh..275....3P} observed a sudden, intense burst of radiation around 14:21 UT, followed by a large-scale, bright, propagating EUV wavefront. 

\begin{figure}[t!]
    \centering
    \includegraphics[width=0.7\textwidth]{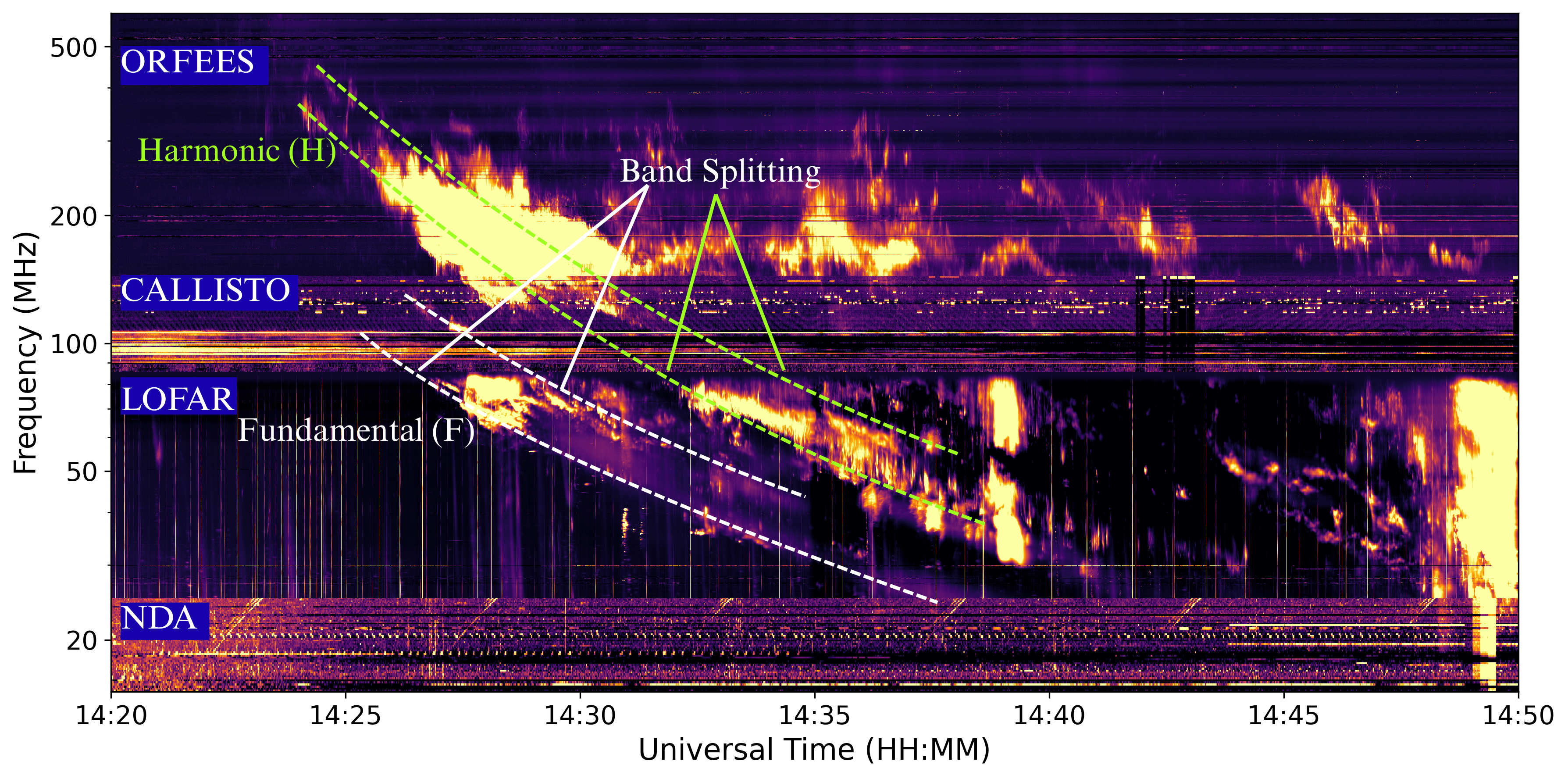}
    \caption{Combined dynamic spectra of the type II radio burst observed with NDA (France), LOFAR (Netherlands), e-Callisto (BIR) and ORFEES (France) solar radio spectrographs. The dashed lines on the spectra shows the fundamental (white) and harmonic (green) band pairs. The band splitting of the fundamental and harmonic bands is also shown. The enhanced emission after type II is a moving type IV radio burst.}
    \label{fig:figure1}
\end{figure}

\begin{figure}[t!]
    \centering
    \includegraphics[width=0.525\textwidth]{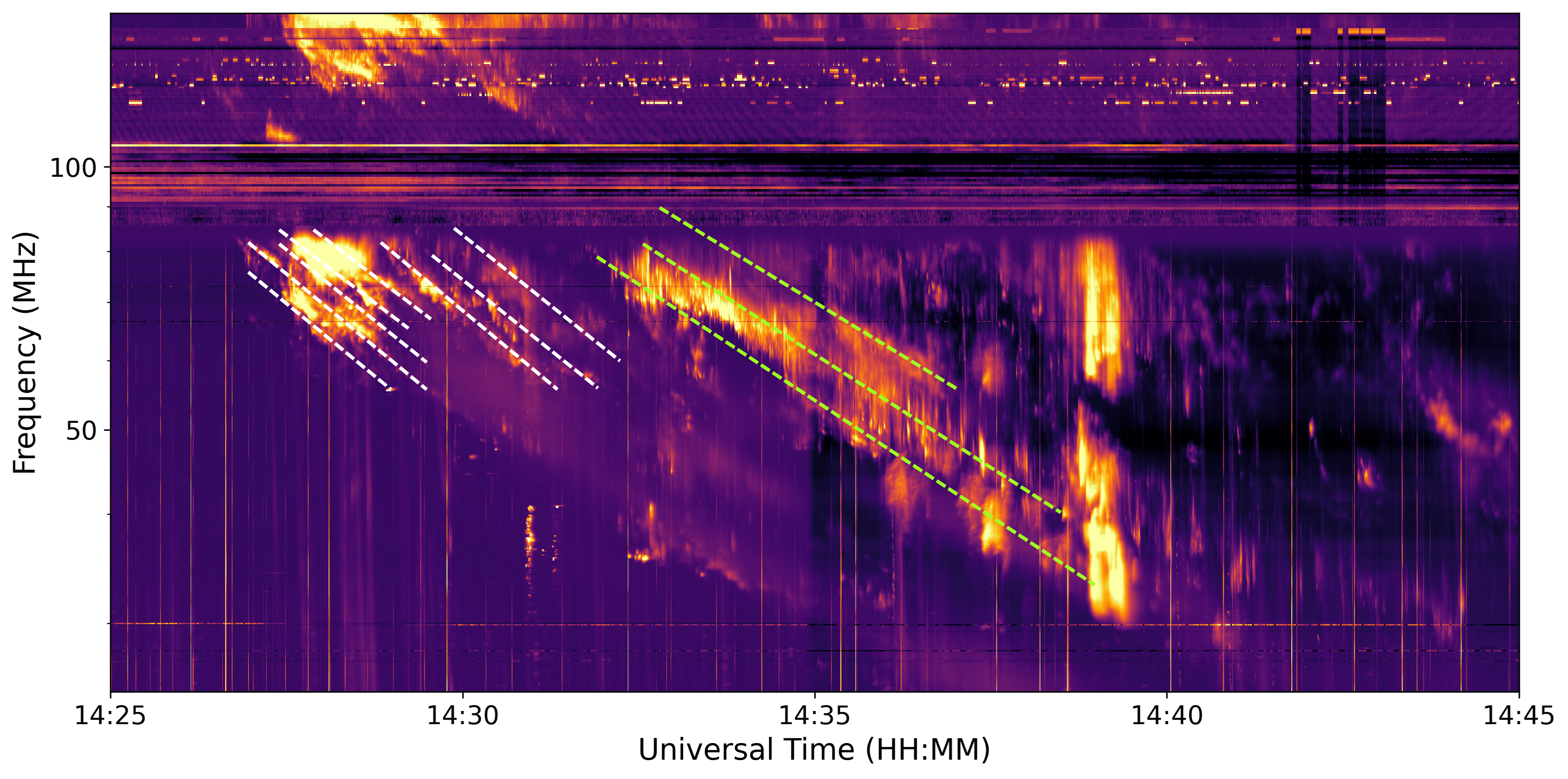}
    \includegraphics[width=0.43\textwidth]{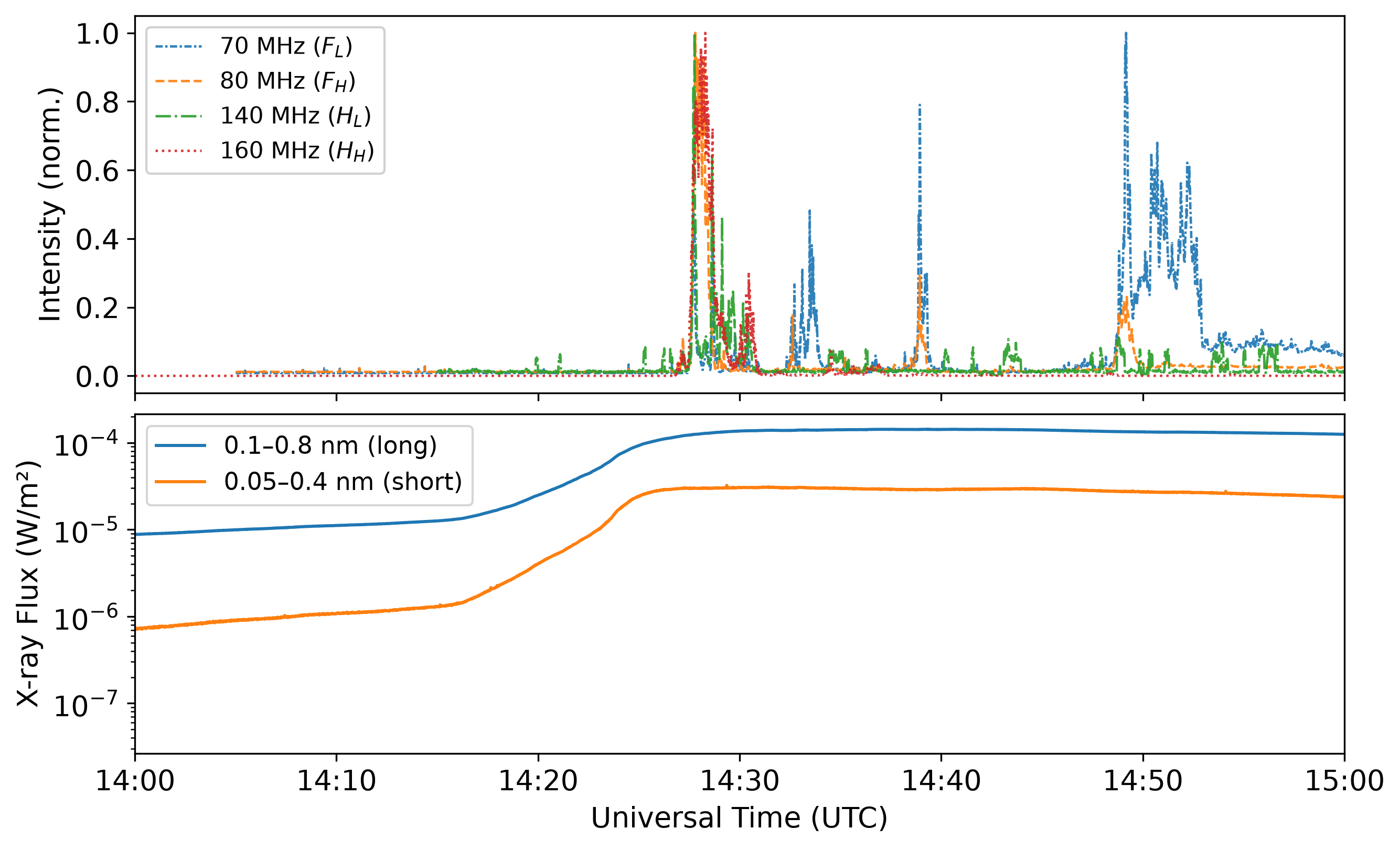}
    \caption{\textbf{Left panel:} The zoomed-in spectra of the type II radio burst observed within the time 14:25 to 14:45 UT and frequency range of 25 to 150 MHz are shown. The dashed lines on the spectra show the multilanes within the fundamental (white) and harmonic (green) pair. \textbf{Right panel:} \textbf{(Top)} Time profile of type II bursts at multiple frequencies. The frequencies are selected as $F_L, F_H, H_L, H_H$ to show the temporal variation of splitbands of the FH pair.
    \textbf{(Bottom)} X-ray flux measured by the GOES-16 satellite in the 0.5--4~\AA\ (orange) and 1--8~\AA\ (blue) channels. The vertical dashed lines mark the flare onset, peak, and end times.}
    \label{fig:figure2}
\end{figure}

A type II solar radio burst was observed with multiple ground-based solar radio spectrographs. Figure \ref{fig:figure1} shows the combined dynamic spectrum observed by the Nançay Decameter Array \citep[NDA;][]{2000GMS...119..321L}, LOw-Frequency ARray \citep[LOFAR;][]{Haarlem2013}, the Compound Astronomical Low-cost Low-frequency Instrument for Spectroscopy and Transportable Observatory \citep[e-CALLISTO;][]{2005SoPh..226..143B} network recorded with the e-Callisto \footnote{\url{https://soleil.i4ds.ch/solarradio/data/2002-20yy_Callisto/2024/05/29/}} Bir Station and the Observations Radiospectrographiques pour FEDOME et l'Etude des Eruptions Solaires \citep[ORFEES;][]{2021JSWSC..11...57H}. The combined dynamic spectra is also marked with the traditional FH and SB pairs. The type II bursts started at 14:24 UT at 450 MHz and ended at 14:43 UT, with the lowest frequency of ~25 MHz. The start timing of the type II radio burst coincides with the peak time of the X-ray flare. The duration of the radio burst is approximately 20 minutes. We also had radio imaging observations from  Nançay radioheliograph \footnote{\url{https://secchirh.obspm.fr/spip.php?page=survey&hour=day&survey_type=1&dayofyear=20240529}} \citep[NRH;][]{1997LNP...483..192K} at multiple frequencies for both FH and SB pairs. We note that, in addition to the FH and SB pairs, there were multiple other visible lanes, as shown in the left panel of figure \ref{fig:figure2}. Table \ref{table1} shows the event timeline as observed by different instruments in various wavelengths.

\begin{table}[t!]
\caption{The timeline of the 29 May 2024 event as seen in different instruments.}
\begin{tabular}{lcccc}
\hline
Instrument       & Start Time & End Time & Particulars \\
\hline
GOES     & 14:11 UT   & 14:55 UT  & X1.4 flare \\
RADIO & 14:24 UT   &     14:43 UT     & Multi-lane type II burst \\
SDO-AIA          & 14:21 UT   &    --       & EUV brightenings \& wavefront \\
STEREO-A EUVI    & 14:22 UT   &        --   & Coronal loop  \\
STEREO-A COR1     & 14:31 UT   &      --     & First appearance of whitelight CME \\
STEREO-A COR2    & 14:53 UT
   &      --     & CME extended view \\
SOHO-LASCO C2    & 14:51 UT   &         --  & CME extended view \\
SOHO-LASCO-C3    & 15:18 UT   &         --  & CME extended view \\
\hline
\end{tabular}
\label{table1}
\end{table}

\section{Data analysis and results} 
      \label{section-sec3}   

To understand the origin of the multilane structures observed in the type II radio burst, a multi-wavelength and multi-instrument imaging approach was employed. AIA (SDO) and EUVI (STEREO-A) captured the low corona, COR1/2 (STEREO-A) the intermediate heights, and LASCO-C2/3 \citep[SOHO;][]{2012SoPh..275....3P} the higher altitudes. 

\subsection{Multi-lane type II burst } 
  \label{section-sec3.1}
Figure \ref{fig:figure1} shows a well-defined type II burst, visible from approximately 14:24 UT to 14:43 UT, exhibiting the slow frequency drift of approximately 0.4 MHz/s on average. The type II burst had clear signatures of fundamental and harmonic lanes, each exhibiting split bands. However, from the zoomed-in dynamic spectra (see figure \ref{fig:figure2}a), it appeared that the split bands are further splitting (marked in figure \ref{fig:figure2}a), suggesting the presence of a multilane structure. As a preliminary approach, the emission lanes (including fundamental and harmonic split bands) were manually traced to obtain a frequency-time relation. The band pairs that appear as fundamental $f_{F}$  and harmonic $f_{H}$ components follow the classical $f_{H}$ = 2$f_{F}$ relation. The splitbands also do satisfy the typical $1:<2$ ratio. The time profiles extracted from ORFEES, CALLISTO, and LOFAR spectra at 70 MHz ($F_{L}$), 80 MHz ($F_{H}$), 140 MHz ($H_{L}$), and 160 MHz($H_{H}$) are shown in the top panel of Figure \ref{fig:figure2}b. Multiple peaks were observed within the same time frame as type II, thereby reaffirming the existence of multilanes. Figure \ref{fig:figure2}b bottom panel shows the corresponding GOES-16 X-ray time profiles in short and long wavelengths.\\

To assess whether the individual emission lanes correspond to shock regions with different compression strengths, we computed the band-split width, density compression ratio, and Alfv\'enic Mach number for each of the three lane pairs. For each pair, co-temporal lower-frequency ($f_{\rm L}$) and
upper-frequency ($f_{\rm U}$) measurements were selected from the dynamic spectrum. Following \cite{Smerd1974} and \cite{Vrsnak2002}, the relative band-split width (BDW) is defined as

\begin{equation}
\boldsymbol{{\rm BDW}
=
\frac{f_{\rm U}-f_{\rm L}}{f_{\rm L}}},
\label{eq:bdw}
\end{equation}

while the corresponding density compression ratio is

\begin{equation}
\boldsymbol{X
=
\left(
\frac{f_{\rm U}}{f_{\rm L}}
\right)^2} ,
\label{eq:compression}
\end{equation}

Assuming a quasi-perpendicular shock geometry ,the Alfv\'enic Mach number is given by

\begin{equation}
\boldsymbol{M_{\rm A}
=
\sqrt{
\frac{X(X+5)}
     {2(4-X)}
     }}
\label{eq:mach}
\end{equation}

The co-temporal frequency pairs used in the analysis were
$(f_{\rm L},f_{\rm U})=(63,83)$ MHz for Pair~A,
$(64,78)$ MHz for Pair~B, and
$(67,82)$ MHz for Pair~C. The derived parameters are
summarized in Table~\ref{table2}.
\begin{table}[t]
\centering
\caption{Band-split parameters derived for the three lane pairs.}
\label{table2}
\begin{tabular*}{\textwidth}{@{\extracolsep{\fill}}cccccc}
\hline
\textbf{Pair} & $\boldsymbol{f_{\rm L}}$ \textbf{(MHz)} & $\boldsymbol{f_{\rm U}}$ \textbf{(MHz)} & \textbf{BDW} & $\boldsymbol{X}$ & $\boldsymbol{M_{\rm A}}$ \\
\hline
\textbf{A} & \textbf{63} & \textbf{83} & \textbf{0.32} & \textbf{1.75} & \textbf{1.62} \\
\textbf{B} & \textbf{64} & \textbf{78} & \textbf{0.24} & \textbf{1.53} & \textbf{1.42} \\
\textbf{C} & \textbf{67} & \textbf{82} & \textbf{0.22} & \textbf{1.50} & \textbf{1.39} \\
\hline
\end{tabular*}
\end{table}

Pair~A yields the largest band-split width, compression ratio, and Mach number, exceeding Pairs~B and C by approximately  14--17\% in $X$, while Pairs~B and C show mutually comparable 
values. This systematic difference indicates that the emission  lanes sample shock segments with different compression strengths.

\subsection{Radio imaging observations} 
  \label{section-sec3.2}
The NRH radio imaging observations were used to track the evolution of the radio emission sources associated with the type II burst. The data spans the frequency range from 150 to 450 MHz.The whole evolution of the source centroids was traced from the preflare to the decay phase at 150.9 MHz, 173.2 MHz, 228 MHz, 270 MHz, 298.7 MHz, 327 MHz, 408 MHz, and 432 MHz, allowing us to identify the spatio-temporal behavior of the radio source as well as, imaging at multiple frequencies allows us to trace the shock propagation through different density layers of the corona. The centroids initially appeared at higher frequencies and, as time progressed, at lower frequencies, consistent with our understanding that radio emissions at higher frequencies originate from regions of higher electron density, located at lower altitudes in the solar corona. The radio imaging further showed that the initial radio emission gradually splits into distinct locations along the shock front over time. When the spectrum shows a multilane structure, we observed a clear separation of imaged radio sources, specifically a double-source structure corresponding to the leading edge and the southern flank of the shock front.

\begin{figure}[t!]
    \centering
        \includegraphics[width=1\textwidth]{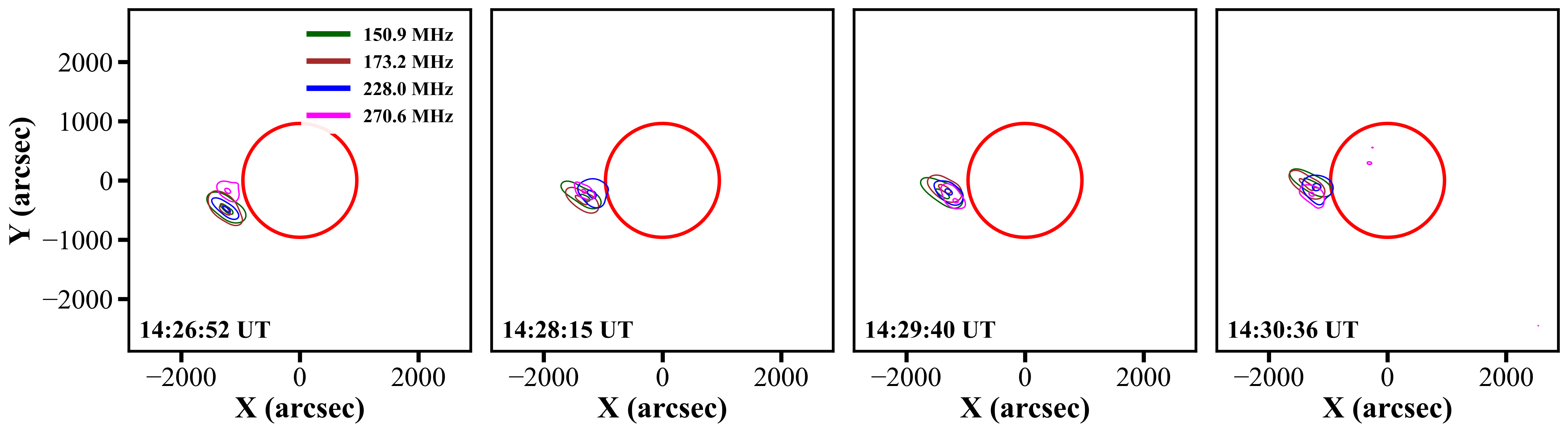}
        \includegraphics[width=1\textwidth]{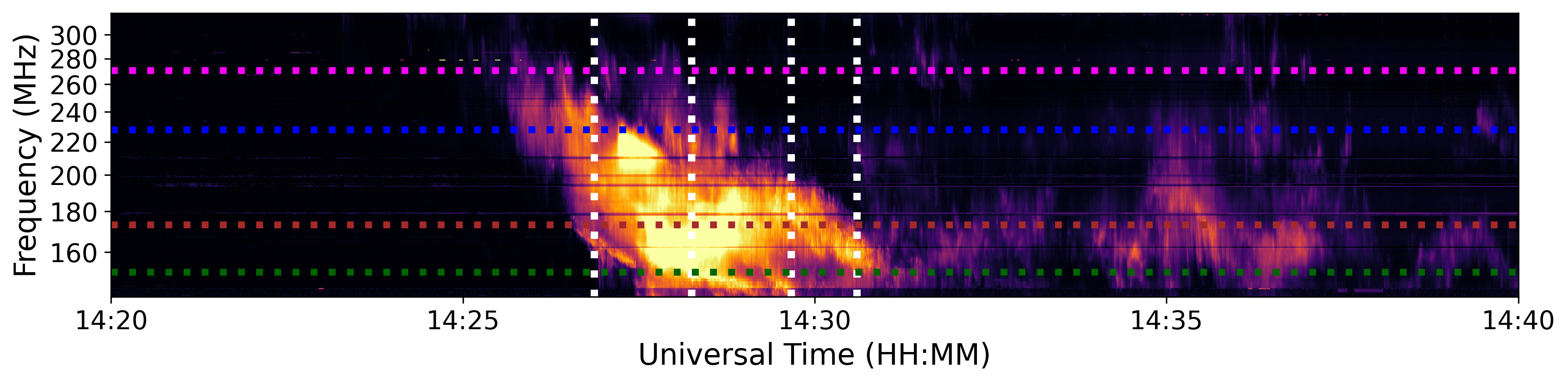}
    \caption{Evolution of radio sources at four frequencies, 150.9 MHz (dark green), 173.2 MHz (brown), 228 MHz (blue), and 270.6 MHz (magenta). The red circle denotes the photospheric Sun. The radio spectra at the bottom are zoomed in in time and frequency, and the dotted lines and colors correspond to frequency and time of the contours made. All the contours are made at $\approx 50 \%$ and $\approx 90 \%$levels.}
    \label{fig:figure3}
\end{figure}

A basic principle in plasma emission theory is that the emitted radio frequency is directly related to the local plasma frequency $f_{p}$, which itself depends on the electron density $n_{e}$ of the corona:
\begin{equation}
    f_{p} \approx 9 \times \sqrt{n_{e}} \quad \text{(in kHz)}
    \label{equation1}
\end{equation}

This indicates that the higher-frequency radio emission should come from higher-density regions, i.e., lower coronal heights. Moreover, conversely, the lower-frequency radio emission should come from lower-density regions, implying a higher corona. But in figure \ref{fig:figure3} we can clearly see that at 14:28:15 UT, the source at $228$~MHz appears at a lower coronal height than the source at $270.6$~MHz, which contradicts Equation \ref{equation1}. Similar inconsistencies were also observed at other time stamps and frequency pairs.

\subsection{EUV and Whitelight observations}
  \label{section-sec3.3}

The EUV eruption and wavefront were clearly observed in all channels of the SDO-AIA and STEREO-EUVI instruments. Top panel of the figure \ref{fig:figure5} shows the time evolution of the EUV wavefront in the SDO-AIA 211 {\AA}  channel and STEREO-EUVI 195 {\AA}  channel. The EUV wavefront, which steepens into a CME as seen in the white-light coronograph images, is marked with arrows in all panels. The STEREO/cor1 and cor2 and SOHO/LASCO-c2 and c3 (see figure \ref{fig:figure5}) show a clear 3-part CME structure. The marked arrows show the leading edge of the shock. The estimated plane-of-sky speed of the CME was approximately 880 km/s (see CDAW \footnote{\url{https://cdaw.gsfc.nasa.gov/CME_list/daily_movies/2024/05/29/}} ). 

\begin{figure}[t!]
    \centering
        \includegraphics[width=1\textwidth]{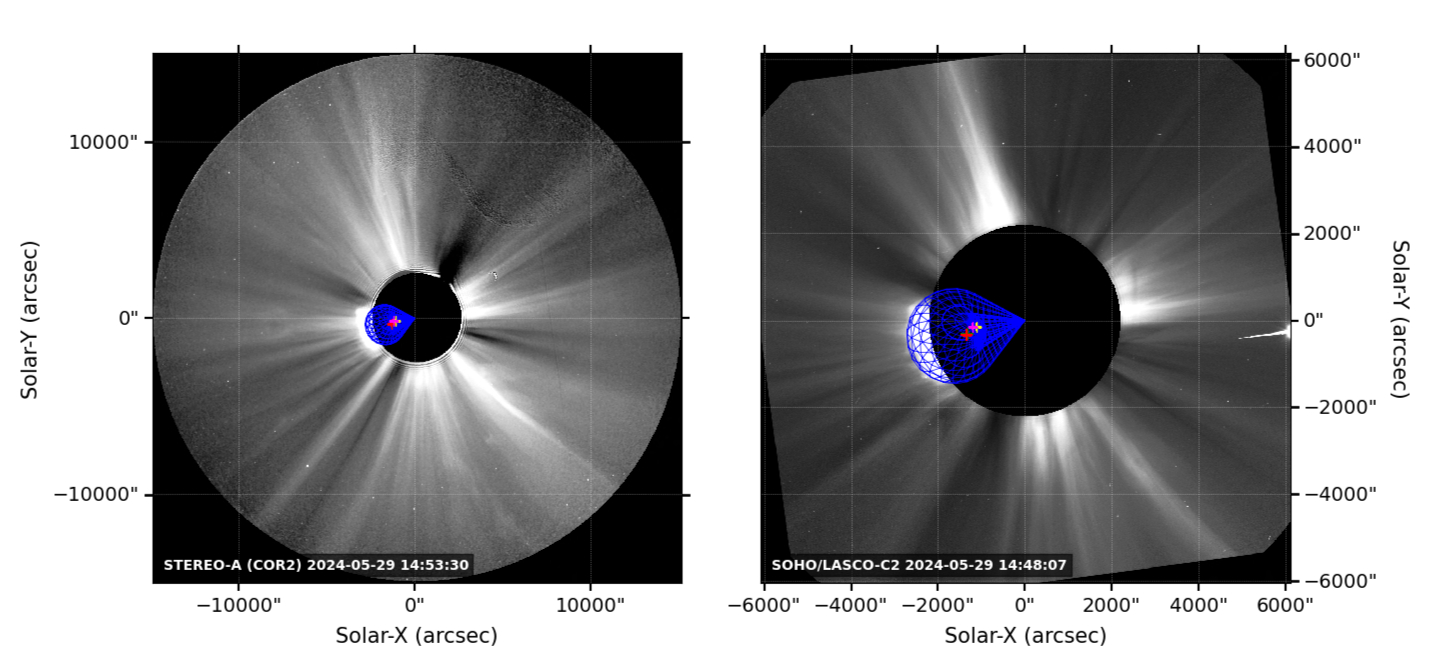}
    \caption{Graduated Cylindrical Shell (GCS) model fitting of the CME observed on 29 May 2024 at 14:48:07 UT using STEREO-A/COR2 (left panel) and SOHO/LASCO-C2 (right panel) coronagraph images. The blue mesh shows the best-fit Graduated Cylindrical Shell flux-rope model. Colored symbols mark the NRH 99\% centriod positions (with error bars) at 150.9 (red), 173.2 (green), 228.0 (yellow), and 270.0 MHz (meganta), onto the coronagraph images. These radio source positions illustrate the radio emission relative to the CME-flux rope structure.    }
    \label{fig:figure4}
\end{figure}

\begin{figure}[t!]
    \centering
    \includegraphics[width=0.93\linewidth]{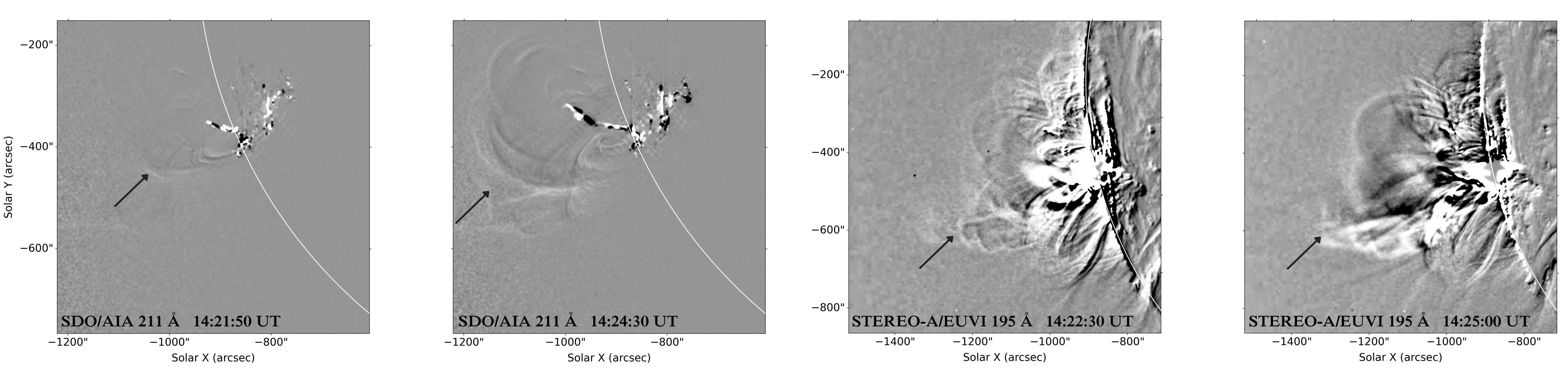}  \includegraphics[width=0.9\linewidth]{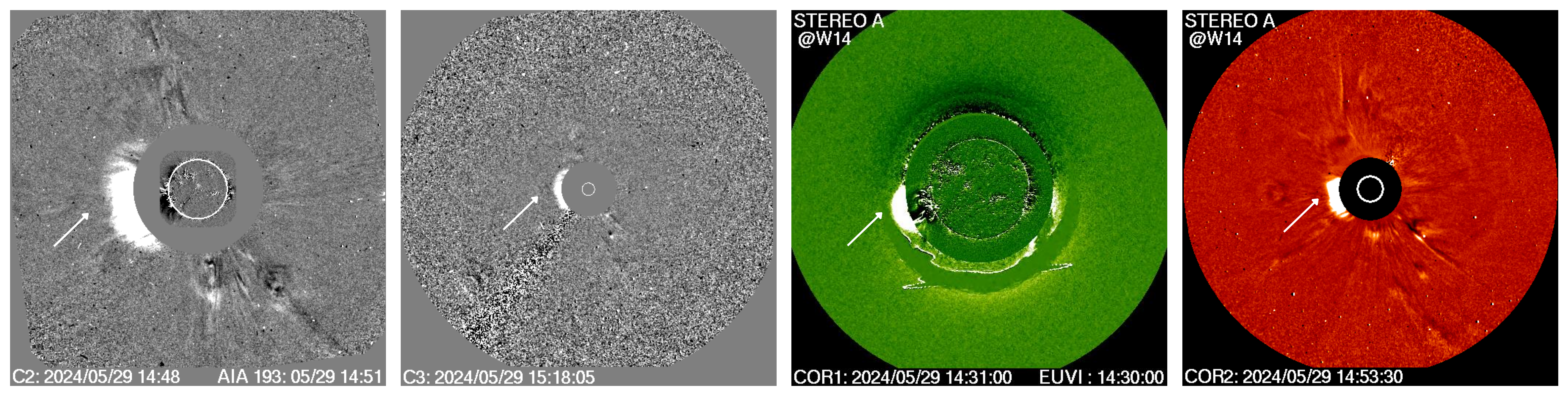}
    \caption{\textbf{Top panel:} The evolution of EUV wave is shown using the snapshots of difference images of the EUV wave seen in SDO/AIA 211 \AA~channel and STEREO-A/EUVI 195 \AA ~channel. \textbf{Bottom panel:} The CMEs first appearance in different coronagraphs' FOV. The two left panels show CME in SOHO/LASCO C2 and C3 coronagraphs. The two right panels show CME in the STEREO-A/SECCHI COR1 and COR2 images.}
    \label{fig:figure5}
\end{figure}

We also incorporated the Graduated Cylindrical Shell (GCS) model to perform a three-dimensional reconstruction of the CME  (Figure \ref{fig:figure4}). The reconstructed CME structure is represented by a blue wireframe mesh. In addition, the NRH 99\% centriod positions at frequencies (150.9 (red), 173.2 (green), 228.0 (yellow), and 270.0 MHz (magenta)) are overlaid on STEREO-A/COR2 and SOHO/LASCO-C2 coronagraph images. The corresponding error bars indicate the positional uncertainties in the radio source centriods, which are determined based on the beam size of the NRH at the respective observed frequencies. This combined analysis enables us to investigate the spatial distribution of radio sources and CME structure in three dimensions, thereby minimizing projection effects and providing a more accurate determination of the radio source locations.

\subsection{CME kinematics}
  \label{section-sec3.4}

\begin{figure}[t!]
    \centering
 \includegraphics[width=0.45\textwidth]{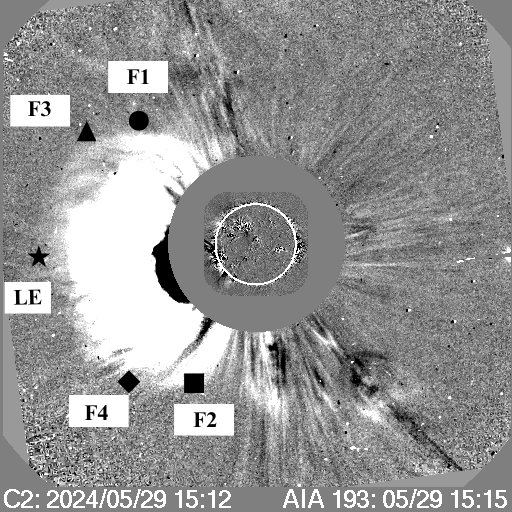}
 \includegraphics[width=0.45\textwidth]{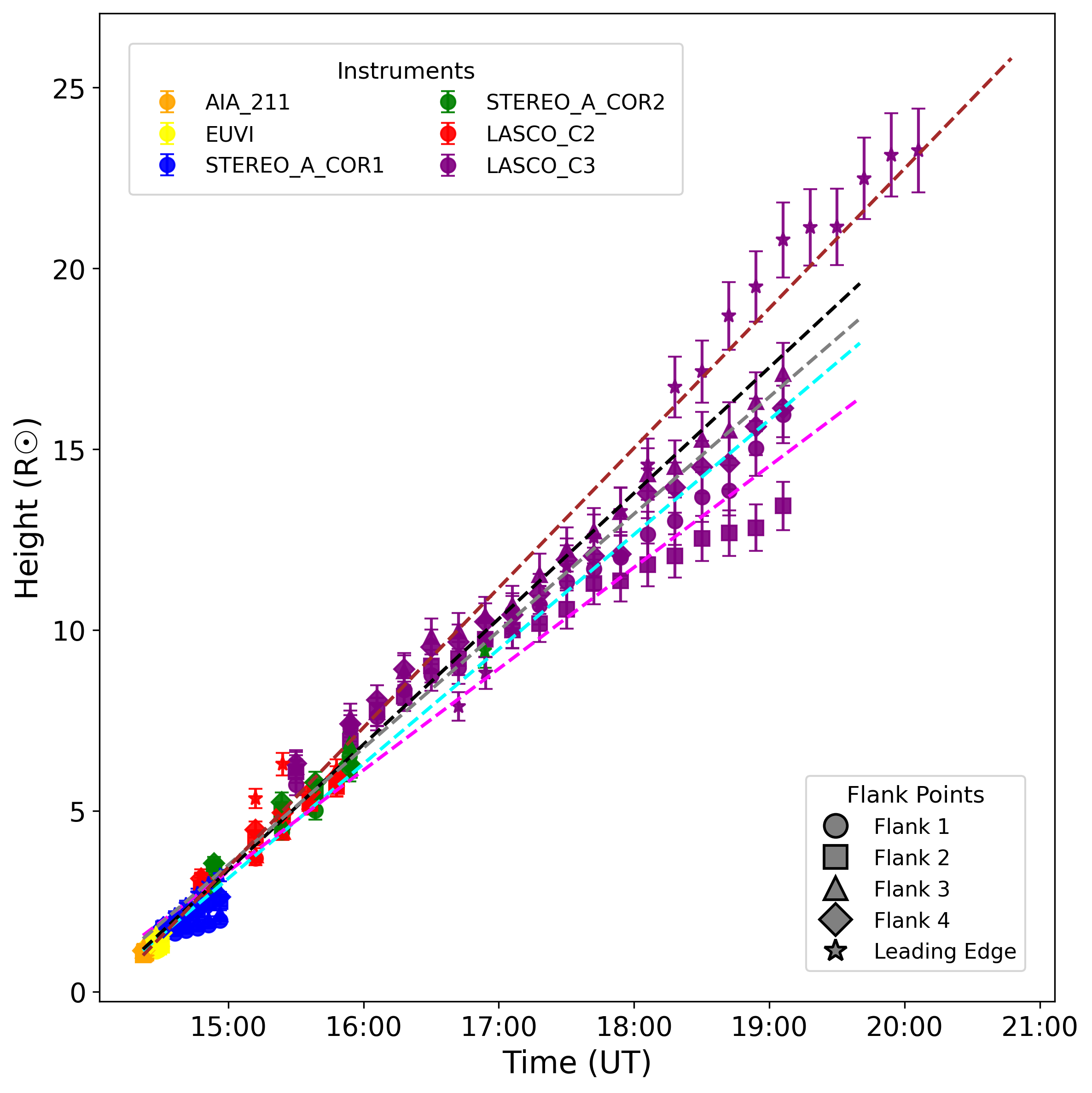}
    \caption{\textbf{Left panel:} Data points marked over the  CME feature at one specific time in the LASCO C2 field of view illustrate how different parts of the CME were traced from the inner to the middle corona. \textbf{Right panel:} Height–time evolution of the CME from the near-Sun region (SDO/AIA) to the middle corona (SOHO/LASCO C3) with measurements at five different structural locations of the CME tracked using multiple instruments (AIA 211, EUVI, LASCO C2, LASCO C3, STEREO-A COR1, STEREO-A COR2). The average speed of the CME is $\approx 880$\, km\,s$^{-1}$. The maximum speed was measured at the CME front ($\approx 1000$\, km\,s$^{-1}$).}
    \label{fig:figure6}
\end{figure}

 In Figure \ref{fig:figure6}, the left panel illustrates the expansion of the CME into the solar atmosphere. To estimate propagation speeds across different parts of the CME structure, we selected five distinct points: one at the leading edge and two at each of the top and bottom flanks (see markings on the figure \ref{fig:figure6}). This selection allows us to demonstrate that different parts of the CME propagate at different speeds. Each point was tracked using six different instruments: AIA 211, EUVI, LASCO C2, LASCO C3, STEREO-A COR1, STEREO-A COR2. The right panel of \ref{fig:figure6} presents the corresponding height-time plots for each selecting point, using observations from these instruments. The slope of each height-time plot provides the speed of the respective CME component, ranging from 800 to 1000 km/s. Thus, it indicates that the lateral expansion rate differed from the radial movement rate of the CME.  This also implies that, at the same time, the flanks were passing through a denser region than the leading edge, because the leading edge is always at a higher height than the flanks. Thus, different type II panels were passing through different density regions in the inhomogeneities/disturbed corona.

\section{Discussion and conclusion} 
      \label{section-sec4}   
The zoomed-in dynamic spectrum Figure (\ref{fig:figure2}) reveals that the type II burst is exhibiting lane structures beyond the classical fundamental-harmonic (FH) and split-band (SB) pairs. The split bands of both the fundamental and harmonic show further splitting, resulting in at least four distinct emission lanes (L1–L4) that are individually traceable in frequency-time space. The band-split analysis per lane pair from fundamental band (Table~\ref{table2}) provides a direct quantitative test of whether the lanes correspond to physically distinct shock segments. Pair~A 
carries a measurably larger density compression ratio and Alfv\'{e}nic Mach number than Pairs~B and C, which are mutually comparable. This directly indicates that the lanes correspond to physically distinct shock segments. A single uniformly compressed shock front propagating through a smoothly stratified corona would have produced a single stable band-split value across all lanes, whereas the observed spread in $X$ from 1.5 to 1.75 is inconsistent with this 
expectation. From the dynamic spectrum alone, three physical scenarios can account for the observed multi-lane configuration: emission from a single shock front propagating through a structured, inhomogeneous corona; emission from several simultaneously active and  spatially distinct shock segments along the CME front; or a combination of both. The band-split difference between Pair~A and Pairs~B and C favours the second or third scenario, since distinct compression ratios implies to distinct upstream plasma conditions. The spatial identification of these segments requires radio imaging.\\

\begin{figure}[t!]
    \centering
 \includegraphics[width=1\textwidth]{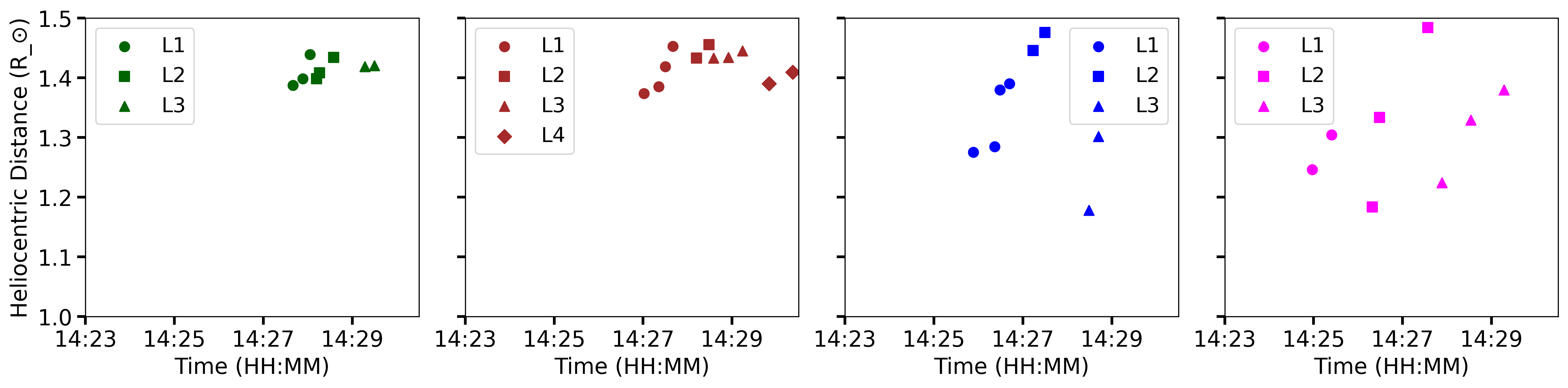}
 \includegraphics[width=1\textwidth]{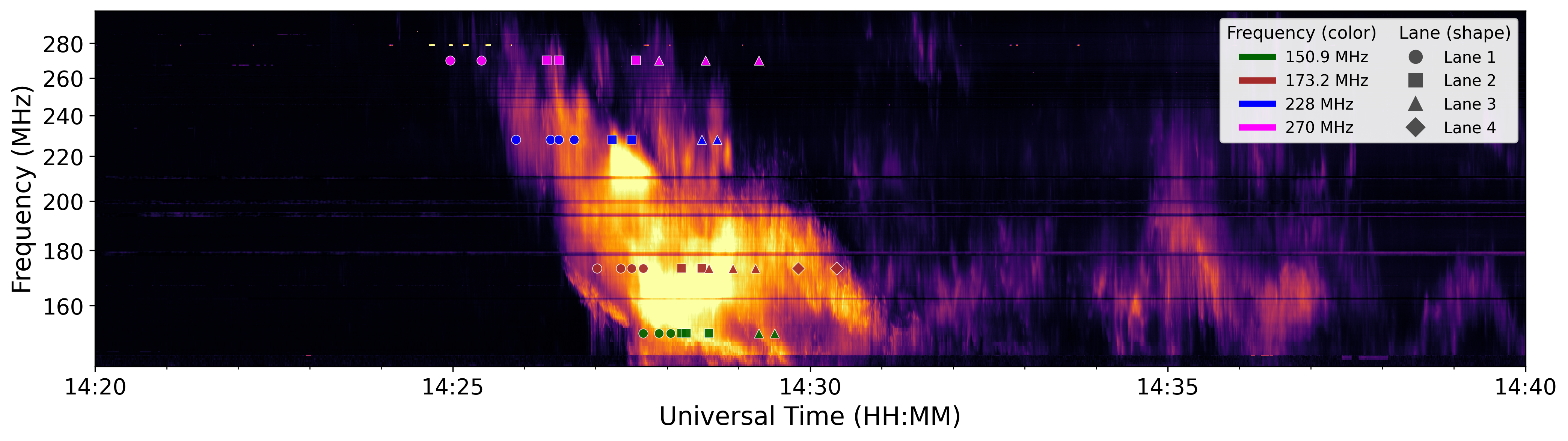}
    \caption{Height--time distribution of radio source centroids identified across multiple lanes and frequencies. \textbf{(Top)} Each point represents the maximum intensity location at a given time and lane, color-coded by frequency and symbol-coded by lane (L1--L4). \textbf{(Bottom)} Dynamic spectrum showing the spectral evolution and multilane structure corresponding to the imaging data. The plot highlights that some lower-frequency sources (e.g., 228 MHz, Lane 3) appear at lower heights than higher-frequency counterparts (e.g., 270.6 MHz, Lane 3), further confirming the frequency--height inversion.}
    \label{fig:figure7}
\end{figure}

The four colors in the plot (Figure \ref{fig:figure3}) represent the four frequencies at which radio imaging was performed (150.9 MHz, 173 MHz, 228 MHz, and 270 MHz). This frequency-dependent imaging was carried out to estimate the spatial locations of the radio sources at different times during the evolution of the type II burst. Since the emission frequency is directly related to the local plasma density and, therefore, to coronal height, imaging at multiple frequencies allows us to trace the shock's propagation through different density layers of the corona.
At each frequency, we identified three to four distinct emission lanes. Each lane corresponds to a different time of occurrence and therefore represents a different height along the shock. Consequently, for a fixed frequency, as expected, the source position evolves with time, reflecting the outward propagation of the shock through regions of varying plasma density. The evolution of the radio counters is shown in Figure \ref{fig:figure7}. However, we found that at some times, higher-frequency emission originated from higher altitudes, contradicting plasma emission theory. We examined the corresponding dynamic spectrum at these particular times. From the dynamic spectra, it is evident that this phenomenon does not occur in individual lanes; it appears only when we track different lanes at different frequencies. To assess whether this inversion is due to the projection effects arising from the event geometry at S20E66, we applied the Graduated Cylindrical Shell (GCS) model to simultaneous SOHO/LASCO-C2 and STEREO-A/COR2 coronagraph images (Figure~\ref{fig:figure4}). Even after accounting
for the three-dimensional geometry, the frequency–height inversion persists, indicating that it is a physical property of the shock rather than a projection artifact.
This is one possible scenario that could explain this peculiar behavior of radio sources. When radio emission arises from multiple shock locations and the shocks pass through an inhomogeneous plasma, they strike in a non-uniform manner. 

The combined spectral and imaging analysis of the multi-lane type II radio emissions reveals a highly structured, dynamically evolving shock propagating in the corona. The multi-frequency imaging observations from NRH show the relation between emission frequency and coronal height. The radio emission locations also strongly suggest that although their emissions are fundamentally governed by the plasma frequency–density relation, they can be significantly influenced by local shock properties and coronal inhomogeneities.

The type II burst at 270 MHz, which was seen at 14:28:30 UT, comes from a coronal region that is denser than the other regions. This is probably because it is located along compressed magnetic structures or higher coronal loops, as seen in the NRH radio pictures (see figure \ref{fig:figure3}). This means that the plasma is being compressed more and that there may be a stronger shock region nearby. The 150 MHz type II burst at 14:27:02 UT, on the other hand, comes from a less dense coronal region at higher altitudes, which fits with the normal density stratification of the solar corona. We also made radio maps of the type IV emission that comes after the type II burst, which was seen at 14:40 UT. Type IV bursts are non-thermal emissions and indicate electrons trapped in post-flare loops or CME structures. These bursts are linked to magnetic restructuring and last longer and cover a wider range of frequencies. The next type III explosion at 14:50 UT shows that intense electron beams are fleeing via open magnetic field lines into space between planets. The presence of type III bursts alongside type II bursts signifies the simultaneous existence of many acceleration regimes during the eruption. 


The larger compression ratio and Alfv\'{e}nic Mach number derived for Pair A suggests that different emission lanes sample shock regions with different local shock parameters rather than a uniform propagating shock. While a higher speed of propagation near the CME leading edge can lead to an enhanced compression, the compression ratio is fundamentally determined by the local Alfv\'{e}nic Mach number, which is a function of both the shock speed and the upstream Alfv\'{e}nic speed. The magnetic field strength and plasma density are different at different locations of the curved CME driven shock, so it is natural that the Mach numbers and shock obliquities $(\theta_{Bn})$ are different, as well. This results in localised differences in the electron acceleration efficiency and plasma compression.

More recently \cite{jebaraj2026} presented theoretical work further showing that turbulence-driven corrugation of collisionless fast-mode shocks naturally generates spatially varying compression, drift velocities and coherence lengths along the shock surface. Therefore, multiple radio lanes may not be the result of discrete shocks, but may be due to simultaneous emission from different corrugated shock segments with different local plasma conditions. Our NRH imaging reveals spatially separated radio sources and measurable difference in compression ratio between the observed pair of lanes, supporting this interpretation. Present observations cannot distinguish between shock corrugation and large-scale coronal inhomogeneity, but they provide strong evidence that the multilane emission is related to spatial variations in the local shock geometry and Mach number rather than variations in propagation speed alone.
This localized density increase explains why the L3 lane at 270 MHz, which has a higher plasma frequency and thus denser plasma, seems higher than the L3 lane at 228 MHz at the same time, which is the opposite of what we would expect. The L3 lane at 228 MHz is lower than the 150 MHz lane, which is in line with the conventional radial density stratification where plasma frequency is directly connected to electron density ($f_p \propto \sqrt{n_e}$) and inversely related to height in the corona. This shows the importance of shock geometry and coronal plasma inhomogeneity in determining the source heights of radio emissions and their radio frequencies. The variations from a straightforward monotonic frequency–height connection suggest that the shock front is not advancing through a homogeneous material. Changes in compression ratio, interactions with already-existing coronal structures (such as streamers or density gradients), and variances in shock obliquity throughout the front all cause local density variation. Consequently, the conventional understanding of fundamental–harmonic (F/H) pairings and split-band (SB) structures, which are often treated as indicators of upstream–downstream density differences, might be insufficient in such complex, high-dynamic situations. The existence of multilane bands probably results from emissions originating from various, specific parts of the shock surface, each reflecting diverse ambient plasma conditions and compression levels , which remain consistent with the results reported by \cite{2020A&A...639A..56J, 2025A&A...703A.271Z}. The observations are consistent with a CME-driven shock whose local compression varies across the shock surface due to variations in Alfv\'en Mach number, shock obliquity and possibly turbulence-driven shock corrugation. The local differences result in several emitting shock segments simultaneously, which explains the observed multilane morphology.

The main conclusions of this study can be summarized as follows:
\begin{itemize}
    \item  Multi-frequency radio imaging suggests that type II burst lanes don't always follow the straightforward monotonic density-height relation that classic coronal models indicate they should.
  \item  These multilane structures are indications that the radio emission is generated by different parts of the CME-driven shock front simultaneously.
  \item  The observed multilane structures indicate that various parts of the CME-driven shock front simultaneously contribute to the radio emission.
\end{itemize}
These results suggest that the observed frequencies and source heights are considerably influenced by shock geometry, plasma inhomogeneity, and evolving CME kinematics. The present observations add a
metric-band imaging case to this growing corpus, reinforcing the conclusion of \cite{2025A&A...703A.271Z} and \cite{Morosan_2025} that individual type II lanes
correspond to distinct shock segments characterised by different local plasma
conditions.

\section{Additional statements}
 
\begin{authorcontribution}
NK has carried out the majority of the data analysis and manuscript preparation. DP has assisted NK with data analysis and manuscript writing. AK has conceptualised the project and supervised the study. 
\end{authorcontribution}

\begin{dataavailability}
The data is publicly available on the respective instrument websites. Reduced and analysed data may be provided on reasonable request to the authors. 
\end{dataavailability}

\begin{codeavailability}
The data analysis is done in Python 3.12 using sunpy 4.0, and all the code is publicly available on GitHub: https://github.com/anshusolar/MultiWavelength-Data.
\end{codeavailability}

\begin{ack}
AK acknowledges the ANRF PM ECRG grant. SOHO data are courtesy of a consortium comprising the Naval Research Laboratory (USA), Max-Planck-Institut für Sonnensystemforschung (Germany), Laboratoire d'Astronomie (France), and the University of Birmingham (UK). SOHO is a project of international cooperation between the European Space Agency (ESA) and NASA. The SOHO/LASCO CME catalog is generated and maintained at the CDAW Data Center by NASA and the Catholic University of America in cooperation with the Naval Research Laboratory. SDO/AIA data are courtesy of NASA/SDO and the AIA science team. We acknowledge Paris observatory, ASTRON, e-Callisto site for radio data.
\end{ack}

\begin{ethics}
\begin{conflict}
The authors declare that they have no conflicts of interest. 
\end{conflict}
\end{ethics}

\bibliographystyle{spr-mp-sola}
\bibliography{references}  

\end{document}